\documentclass[12pt]{iopart}
\usepackage{graphicx}
\usepackage{subfigure}
\usepackage{multirow}
\usepackage{booktabs}
\usepackage{url}
\makeatletter
\def\blx@maxline{77}
\makeatother

\usepackage{arydshln}
\expandafter\let\csname equation*\endcsname\relax
\expandafter\let\csname endequation*\endcsname\relax
\usepackage{todonotes}
\usepackage[T1]{fontenc}
\usepackage{amsmath} 

\usepackage{comment}
\usepackage{parskip}

\begin{document}


\title[A quantitative assessment of Geant4 for QA in ion beam therapy]{A quantitative assessment of Geant4 for predicting the yield and distribution of positron-emitting fragments in ion beam therapy}

\author{Andrew~Chacon$^{1}$, Harley~Rutherford$^{1,2}$, Akram~Hamato$^{3}$, Munetaka~Nitta$^{3}$, Fumihiko~Nishikido$^{3}$, Yuma~Iwao$^{3}$, Hideaki~Tashima$^{3}$, Eiji~Yoshida$^{3}$, Go~Akamatsu$^{3}$, Sodai~Takyu$^{3}$, Han Gyu Kang$^{3}$, Daniel~R.~Franklin$^{4}$, Katia~Parodi$^{5}$, Taiga~Yamaya$^{3}$, Anatoly~Rosenfeld$^{2,6}$, Susanna~Guatelli$^{2,6}$, Mitra~Safavi-Naeini$^{1,2,7}$}

\address{$^1$ Australian Nuclear Science and Technology Organisation (ANSTO), NSW, Australia}
\address{$^2$ Centre for Medical Radiation Physics, University of Wollongong, Wollongong NSW 2522 Australia}
\address{$^3$ National Institutes for Quantum Science and Technology, Chiba, Japan}
\address{$^4$ School of Electrical and Data Engineering, University of Technology Sydney, Australia}
\address{$^5$ Ludwig-Maximilians-Universit{\"a}t M{\"u}nchen, Department of Medical Physics, Faculty of Physics, Garching b. Munich, Germany}
\address{$^6$ Illawarra Health and Medical Research Institute, University of Wollongong, Wollongong NSW 2522 Australia}
\address{$^7$ Brain and Mind Centre, University of Sydney, Sydney, NSW, Australia}

\ead{mitras@ansto.gov.au}

\maketitle

\begin{abstract}


Purpose: To compare the accuracy with which different hadronic inelastic physics models across ten Geant4 Monte Carlo simulation toolkit versions can predict positron-emitting fragments produced along the beam path during carbon and oxygen ion therapy.

Materials and Methods: Phantoms of polyethylene, gelatin or poly(methyl methacrylate) were irradiated with monoenergetic carbon and oxygen ion beams. Post-irradiation, 4D PET images were acquired and parent $^{11}$C, $^{10}$C and $^{15}$O radionuclides contributions in each voxel were determined from the extracted time activity curves. Next, the experimental configurations were simulated in Geant4 Monte Carlo versions 10.0 to 11.1, with three different fragmentation models - binary ion cascade (BIC), quantum molecular dynamics (QMD) and the Liege intranuclear cascade (INCL++) - 30 model-version combinations. Total positron annihilation and parent isotope production yields predicted by each simulation were compared between simulations and experiments using normalised mean squared error and Pearson cross-correlation coefficient. Finally, we compared the depth of maximum positron annihilation yield and the distal point at which positron yield decreases to 50\% of peak between each model and the experimental results.

Results: Performance varied considerably across versions and models, with no one version/model combination providing the best prediction of all positron-emitting fragments in all evaluated target materials and irradiation conditions. BIC in Geant4 10.2 provided the best overall agreement with experimental results in the largest number of test cases. QMD consistently provided the best estimates of both the depth of peak positron yield (10.4 and 10.6) and the distal 50\%-of-peak point (10.2), while BIC also performed well and INCL generally performed the worst across most Geant4 versions.

Conclusions: Best spatial prediction of annihilation yield and positron-emitting fragment production during carbon and oxygen ion therapy was found to be 10.2.p03 with BIC or QMD. These version/model combinations are recommended for future heavy ion therapy research.

\end{abstract}

\section{Introduction}

One of the chief advantages of particle therapy as a treatment for cancer is the high dose gradient between the treatment area and surrounding regions \cite{Durante2017}. However, to enable the smallest possible margins, particle range must be accurately verified relative to the patient anatomy. Many authors have investigated the post-irradiation imaging of positron-emitting fragments via positron emission tomography (PET) for range verification of the primary beam during particle therapy \cite{Enghardt2004,Parodi2007,Paganetti2012,Kraan2015,Mohammadi_2019,Pei2022,Sun2021,SHIBA2021}, with several prototype in-beam PET scanners currently under active development \cite{Pennazio2018,Grevillot2020,Akamatsu2019,Parodi2022}. Monte Carlo simulation methods, implemented using platforms such as Geant4, are now widely used in particle therapy - applied clinically as a secondary quality assurance (QA) technique to verify dose during particle therapy \cite{geant4,Jia2012,Tseung2013} and there is active research in using it for developing and testing range models \cite{Ferrero2018,Kozowska2019,Zhang2021}, PET image-to-dose models \cite{2019Hofmanna,2019Hofmannb,Rutherford2020,Rutherford2022}, and designing novel PET scanners optimised for range and dose verification \cite{Ahmed2020,Fiorina2018}.


Good models for nuclear fragmentation processes are essential for faithfully simulating imaging applications in particle therapy, such as PET-based dose estimation methods for quality assurance, since the production and distribution of positron-emitting radionuclide fragments directly affects the quality of the resulting image \cite{2019Hofmanna,2019Hofmannb,Rutherford2020,Parodi2018}. Currently, the Geant4 Monte Carlo simulation framework offers a choice of three hadronic inelastic fragmentation models that are appropriate for particle therapy - binary ion cascade (BIC), quantum molecular dynamics (QMD), and Li\`{e}ge intranuclear cascade (INCL++) \cite{geant4,g4physics,incl++}\footnote{INCL++ is considered the most appropriate option for neutron spallation simulations, but is included here for completeness \cite{Boudard2002}.}. In our previous study, we evaluated these models by comparing the spatial distributions of positron-emitting radionuclides predicted following irradiation of PMMA, gelatin and polyethylene targets by monoenergetic carbon and oxygen ion beams (simulated using Geant4 10.2.p03) to equivalent results estimated from experimentally-obtained PET data \cite{Chacon2019}. The BIC model provided the best estimates overall; however, none of the models provided a perfect fit in all evaluated cases, and some significant discrepancies were observed. Other authors have noted similar variations in simulation fidelity in different contexts; most recently, Wro\'{n}ska et\ al. \cite{Wroska2021} evaluated two models, BIC and QBBC, with several versions of Geant4 for prediction of prompt gamma emission during proton therapy with a PMMA target. They concluded that Geant4 10.4p02's BIC implementation with high-precision neutron data libraries and binary ion fragmentation models provided the best agreement with the observed experimental data for this use case, with the simulations performed using other version/model combinations producing features which were absent in the experimental data. Pei et\ al. \cite{Pei2022} evaluated the ability of Geant4 to predict the range of particles using Geant4 version 10.6 and the Bertini fragmentation model. They concluded that Geant4 was able to accurately predict within 1~mm the positron activity peak and distal falling edge using two carbon monoenergetic beams and a single PMMA phantom.

Since the publication of our previous study, there have been several updates to Geant4; specifically, six minor releases (versions 10.x) and one major release (version 11, which has since been updated to version 11.1). Each of these releases includes modifications to the physics models implemented in Geant4, which can affect the simulation of positron-emitting fragment production in particle therapy. 


In this work, we have extended the previous study, and present a quantitative evaluation of Geant4's ability to predict positron-emitting fragment production across a total of ten different stable versions (10.0.p04, 10.1.p03, 10.2.p03, 10.3.p03, 10.4.p03, 10.5.p01, 10.6.p03, 10.7.p02, 11.0 and 11.1) which have followed the previous major release (10.0) for each of the three different fragmentation models \cite{Chacon2019}. 

In addition to the normalised mean squared error (NMSE) metric used in the previous study, three additional metrics - the Pearson cross-correlation coefficient (CCC), the depth of the positron annihilation peak, and the depth at which positron annihilation intensity has decreased to 50\% of peak are also used to compare the shape of the predicted positron-emitting fragment distribution to the experimentally-measured distribution. The experimental absolute yields of the dominant positron-emitting fragment species ($^{10}$C, $^{11}$C and $^{15}$O) created during irradiation of gelatin, PMMA and polyethylene block phantoms with beams of $^{12}$C ions with energies of 148.5, 290.5 and 350~MeV/u and $^{16}$O ions with energies of 148 and 290~MeV/u were obtained across the full width at tenth maximum (FWTM) of the beam from a previous study conducted using the OpenPET scanner at HIMAC, Japan \cite{Akamatsu2019,Chacon2019}. Akamatsu2019 results in each of the three target materials and 5 ion/energy combinations were then compared to those predicted in equivalent simulations performed in Geant4 using each of hadronic fragmentation models (BIC, QMD and INCL++) across the ten evaluated Geant4 versions for a total of 150 unique target/ion/energy/version/model test conditions. The total positron yields and yields of the individual positron-emitting fragment species from each model and Geant4 version were then compared to experimental measurements using each of the chosen metrics at the entrance, build-up and Bragg peak, and tail regions.



\section{Materials and Methods}\label{sec:method}

This section describes the methods used for obtaining and quantitatively comparing the experimental and simulated positron annihilation profiles. The general approach is similar to that used in our previous study (see \cite{Chacon2019}); however, it has been extended to include a much wider range of Geant4 versions, and  additional comparison metrics are introduced.

The experimental methods used to estimate the total positron annihilation profile and activity of the dominant positron-emitting fragment isotopes ($^{11}$C, $^{10}$C and $^{15}$O) are briefly summarised in Section~\ref{sec:method/experiment}. Equivalent simulation configurations were constructed for each Geant4 version under test, and the total positron annihilation profile and activity of $^{11}$C, $^{10}$C and $^{15}$O were predicted for each beam ion/energy, target material, hadronic inelastic fragmentation model and Geant4 version; the design and parameters of these simulations are described in detail in Section~\ref{sec:methods/simulation}. 

The simulated annihilation profiles are compared with the experimental annihilation profiles using the following  metrics in each of the entrance, build-up and Bragg peak and tail regions:

\begin{itemize}
\item Normalised mean squared error (NMSE); and
\item Pearson cross-correlation coefficient (CCC)
\end{itemize}

Additionally, the depth of positron annihilation peak and the depth of the distal point at which the magnitude of the positron annihilation profile decreases to 50\% of the peak value are evaluated. All metrics are described in detail in Section~\ref{sec:method/evaluation}.

\subsection{Experimental configuration}\label{sec:method/experiment}

The experimental data obtained in our 2019 paper were used as the ground truth for this simulation study; a detailed description of the experimental procedures is presented in that paper \cite{Chacon2019}. In summary, phantoms constructed from either pure PMMA, polyethylene or gelatin (encased in a thin-walled PMMA container), each with dimensions of 100~mm$\times$100~mm$\times$300~mm, were irradiated with monoenergetic carbon or oxygen ion beams of various energies - three for carbon ions and two for oxygen (see \tablename~\ref{tbl:beamParam}). Positron annihilation profiles (with respect to depth in the target) were obtained using the whole-body DOI-PET scanner prototype developed at QST \cite{Akamatsu2019}. These profiles were decomposed into the individual population of each of the dominant parent positron-emitting fragments ($^{11}$C, $^{10}$C and $^{15}$O) at $t = 0$ (end of irradiation period) by fitting the observed time-decay curves in each voxel to a multiexponential decay model.

\begin{table}
	\centering
	\caption{Beam parameters for each ion species and energy. The energy spread is 0.2~\% of nominal energy in each case; 95\% confidence intervals are given for beam flux.}
	\begin{tabular}{p{0.06\columnwidth}p{0.25\columnwidth}p{0.15\columnwidth}p{0.15\columnwidth}p{0.25\columnwidth}}
		\toprule
		{\bf Ion}	&{\bf Energy (MeV/u)}	&{\bf $\sigma_x$ (mm)} &{\bf $\sigma_y$ (mm)}   & {\bf Beam flux (pps)}   \\
		\midrule
		$^{12}$C& 	148.5 &2.77 & 2.67 &1.8$\times$10$^{9}\pm 3.8\times$10$^{7}$ \\
		\addlinespace[0.5em]
		$^{12}$C& 	290.5 &3.08 & 4.70&1.8$\times$10$^{9}\pm 6.4\times$10$^{7}$ \\
		\addlinespace[0.5em]
		$^{12}$C& 	350 &2.50 & 2.98&1.8$\times$10$^{9}\pm 4.6\times$10$^{7}$ \\
		\addlinespace[0.5em]
		$^{16}$O& 	148 &2.79 & 2.89&1.1$\times$10$^{9}\pm 2.8\times$10$^{7}$ \\
		\addlinespace[0.5em]
		$^{16}$O& 	290 &2.60 & 4.90&1.1$\times$10$^{9}\pm 7.0\times$10$^{7}$ \\
		\bottomrule
	\end{tabular}
	\label{tbl:beamParam}
\end{table}

\subsection{Simulation parameters}\label{sec:methods/simulation}

The same beam parameters, phantom compositions and geometries used in the experimental measurements were modelled in each version of Geant4. Apart from minor modifications to the simulation source code required due to version-to-version changes in certain Geant4 application programming interfaces (APIs), the code was identical across versions. Simulations were performed using each of the 10 most recent stable releases of Geant4: 10.0.p04, 10.1.p03, 10.2.p03, 10.3.p03, 10.4.p03, 10.5.p01, 10.6.p03, 10.7.p02, 11.0 and 11.1. For brevity, the patch number will be dropped when referring to the version of Geant4.

For each version of Geant4, three alternative hadronic ion fragmentation models were evaluated - Binary Ion Cascade (BIC), Quantum Molecular Dynamics (QMD) and Li\`ege Intranuclear Cascade (INCL) models\footnote{The INCL model was developed specifically for spallation reactions but is included in this study as it can also model fragmentation.} \cite{g4physics,incl++}. All simulations modelled electromagnetic interactions using the standard option 3 list ({\tt G4EmStandardPhysics\_option3}). The remaining physics processes, including hadronic physics models, are listed in Table ~\ref{tbl:physics-list}.

The location of each positron annihilation, as well as the identity of the parent isotope which decayed to emit each positron (principally $^{11}$C, $^{10}$C and $^{15}$O), were scored with a resolution of 1.5~mm$^3$ to match the voxel dimensions of the experimental OpenPET image reconstruction output. The pristine positron annihilation profiles were convolved with a 2.3~mm FWHM Gaussian filter to simulate the point spread function of the PET system.

\subsection{Evaluation methods}\label{sec:method/evaluation}

The irradiated target was divided into three separate regions for analysis since different physics processes dominate in each: the entrance region, the build-up and Bragg peak region, and the tail region. This segmentation is defined in the same way as in our previous paper \cite{Chacon2019}; in summary, the central build-up and Bragg peak region is defined as follows:

\begin{itemize}
	\item The proximal edge in the $z$ dimension (along the path of the beam) is defined as the first point at which the dose deposited along the central axis exceeds the entrance plateau dose by more than 5\% of the difference between peak dose and the entrance plateau dose; and
	\item The distal edge in $z$ is defined as the last point at which the deposited dose is greater than 5\% of the absolute peak dose value.
\end{itemize}

The entrance region is then defined as the region proximal to the build-up and Bragg peak region, while the tail region is defined as the region distal to the build-up and Bragg peak region.

The yields of positron-emitting nuclei are defined via \eqref{eqn:relYields}:

\begin{equation}\label{eqn:relYields}
	\mathrm{Yield~(Isotope)} = \frac{N~(\mathrm{Isotope})}{N~\mathrm{(Primary)}}
\end{equation}

where $N~\mathrm{(Isotope)}$ is the yield of the isotope under study in that region and $N~\mathrm{(Primary)}$ is the total number of primary particles. Yields were calculated in each voxel along the path of the beam.

Three different metrics were chosen to quantify the accuracy of each model in Geant4: the normalised mean squared error (NMSE),  the Pearson cross-correlation coefficient (CCC), and the range (depth along the path of the beam) of both the position annihilation peak and the point beyond the peak at which positron annihilation intensity decreases by 50\%. The NMSE was used to measure the absolute difference between the experimental measurements and simulation-predicted positron yields in each region:

\begin{equation}\label{eqn:NMSE}
	NMSE = \frac{\displaystyle\sum_{i=1}^{N_{reg}}|S_i - E_i|^2}{\displaystyle\sum_{i=1}^{N_{reg}} |E_i|^2}
\end{equation}

where $S_i$ and $E_i$ are the simulation and experimental yields in the $i$th voxel of the $N_{reg}$ voxels in region $reg$ (with a lower value indicating a better match).

The Pearson CCC was used to compare the shapes of the simulation-predicted positron-emitting fragment distributions to the experimental measurements:

\begin{equation}\label{eqn:crossCor}
CCC = \frac{\sum_{i=1}^{N_{reg}}  (S_{norm,i} - \overline{S_{norm}}) (E_i - \overline{E_{norm}})    }{\sqrt{ (\sum_{i=1}^{N_{reg}} (S_{norm,i} - \overline{S_{norm}})^2)  (\sum_{i=1}^{N_{reg}}      (E_{norm,i} - \overline{E_{norm}})^2)}}
\end{equation}

where $S_{norm,i}$ and $E_{norm,i}$ are the normalised simulation and experimental yields in the $i$th voxel of the $N_{reg}$ voxels in region $reg$. Normalisation is performed by dividing each $S_i$ and $E_i$ by the maximum value in its respective region. $\overline{S_{norm}}$ and $\overline{E_{norm}}$ are the mean values in each region.

When comparing the models, the closer the CCC between the model prediction and the experimental estimate of positron-emitting fragment distribution is to +1, the more accurate the prediction. A CCC greater than 0.95 is considered to indicate that the simulation has accurately predicted the shape of the yield.

For each version of Geant4, phantom, beam type and energy, the NMSE and CCC were calculated for both total annihilation photon yield profiles and also for the profiles of the three main positron-emitting fragment species ($^{10}$C, $^{11}$C and $^{15}$O). The calculation was repeated for each of the $N_{reg}$ regions (entrance, build-up and Bragg peak, and tail regions). The NMSE and the CCC were then compared across all evaluated Geant4 versions for each region, phantom material and beam type.

A total of 5 energy/ion combinations are evaluated (carbon ions at three energies and oxygen ions at two energies). For oxygen ions, three target materials (gelatine, PMMA and polyethylene) are evaluated for total positron annihilation yield and $^{11}$C/$^{10}$C/$^{15}$O yield. For carbon ions, the same three target materials are evaluated for total positron annihilation yield and $^{11}$C/$^{10}$C yield and two for $^{15}$O yield (polyethylene is omitted since it is not possible to produce $^{15}O$ fragments with a $^{12}C$ ion beam and a PE target which only contains carbon and hydrogen). Thus, there are a total of 15 cases evaluated for total positron annihilation, $^{11}$C yield and $^{10}$C yield, while there are 12 cases evaluated for $^{15}$O yield.

For range calculations, the difference between the depths at which the positron annihilation yield reached its maximum value in the experiment and simulation was calculated (see \eqref{eqn:range}). Additionally, the point distal to this maximum at which positron annihilation yield decreases to 50\% of the maximum value was also compared between experiment and simulation. For each version and model, the mean differences between the experimental and simulation-based values, as well as the standard deviations and maximum differences were calculated across all test cases (ion species, energies and target materials).

\begin{equation}\label{eqn:range}
	\delta_{voxel} = R_{simulation} - R_{experiment}
\end{equation}

where $R_x$ is the range (depth) of the voxel with the maximum value (or, for distal 50\%-of-peak, the first distal voxel to fall below 50\% of the maximum value) in either the simulation or experiment.

\section{Results and Discussion}\label{sec:resultsDiscussion}

The number of cases in which each version/model combination performed the best or equal-best in terms of each of the evaluated metrics are counted across all simulations in the entrance, build-up and Bragg peak and tail regions, and summarised in this section. Detailed results for each experiment are included in the Supplementary Materials. 

For the NMSE metric, equal best is defined by firstly identifying the version/model with the best mean NMSE, and then accepting any additional version/model combination whose mean NMSE is within the confidence interval of the best-performing version/model as being equal. All confidence intervals are $\pm 2\sigma$ (i.e. 95\% assuming a Gaussian distribution).

\subsection{Entrance region}\label{sec:resultsDiscussion-entrance}

In the entrance region, positron-emitting fragments are created by target fragmentation rather than projectile fragmentation. The projectile ions lose energy via Coulomb interactions, slowing down at an approximately constant rate as they traverse this region, with only gradual changes to projectile/target cross sections. As a result, the positron-emitting fragment distributions are expected to exhibit an approximately flat depthwise profile in this region.

NMSE and Pearson CCC results between simulation and experimental total positron annihilation profiles in the entrance region are summarised in Tables \ref{tbl:NMSE_Entrance_positron} and \ref{tbl:CCC_Entrance_positron}, respectively, with corresponding figures shown in Supplementary Material Section 1.

\begin{table}
\caption{Number of test cases for which each Geant4 version/model combination achieved the lowest or equal-lowest NMSE in the {\bf entrance region}. \textbf{Bold text} denotes the version/model achieving the highest (or equal-highest) number of best results for each combination of ion/energy/target.}
\scriptsize
\centering
\begin{tabular}{p{0.08\columnwidth}p{0.04\columnwidth}p{0.04\columnwidth}p{0.04\columnwidth}p{0mm}p{0.04\columnwidth}p{0.04\columnwidth}p{0.04\columnwidth}p{0mm}p{0.04\columnwidth}p{0.04\columnwidth}p{0.04\columnwidth}p{0mm}p{0.04\columnwidth}p{0.04\columnwidth}p{0.04\columnwidth}} 
\toprule 
\multirow{3}{*}{\bf Version}&\multicolumn{3}{c}{\bf Total}&&\multicolumn{3}{c}{\bf $^{11}$C} &&\multicolumn{3}{c}{\bf $^{10}$C} &&\multicolumn{3}{c}{\bf $^{15}$O}\\
\cmidrule{2-4} \cmidrule{6-8} \cmidrule{10-12} \cmidrule{14-16}
&{\bf BIC}	&{\bf QMD} &{\bf INCL}&&{\bf BIC}	&{\bf QMD} &{\bf INCL}&&{\bf BIC}	&{\bf QMD} &{\bf INCL}&&{\bf BIC}	&{\bf QMD} &{\bf INCL}\\
\midrule 
10	    &\bf 6	&0	&0	&&\bf 11	&3	&2	&&0	&0	&2	&&6	&2	&0	\\ \addlinespace[0.5em]
10.1	&\bf 6	&0	&0	&&\bf 11	&3	&2	&&0	&0	&3	&&6	&2	&0	\\ \addlinespace[0.5em]
10.2	&5	    &0	&0	&&\bf 11	&3	&2	&&0	&0	&3	&&5	&2	&0	\\ \addlinespace[0.5em]
10.3	&\bf 6	&0	&0	&&6	&0	&0	        &&5	&3	&6	&&4	&2	&0	\\ \addlinespace[0.5em]
10.4	&\bf 6	&0	&0	&&1	&0	&0	        &&2	&3	&6	&&\bf 9	&2	&0	\\ \addlinespace[0.5em]
10.5	&2	    &0	&0	&&0	&0	&0	        &&2	&5	&\bf 8	&&3	&1	&0	\\ \addlinespace[0.5em]
10.6	&4	    &0	&0	&&1	&0	&0	        &&0	&0	&5	&&5	&4	&0	\\ \addlinespace[0.5em]
10.7	&3	    &0	&0	&&1	&0	&0	        &&0	&0	&5	&&5	&2	&0	\\ \addlinespace[0.5em]
11	    &4	    &0	&0	&&1	&0	&0	        &&0	&0	&5	&&5	&2	&0	\\ \addlinespace[0.5em]
11.1	&4	    &0	&0	&&0	&0	&0	        &&0	&0	&5	&&5	&2	&0	\\ 
\bottomrule
\hline
\end{tabular}
\label{tbl:NMSE_Entrance_positron}
\end{table}

For the entrance region, the BIC model implemented in Geant4 versions 10, 10.1, 10.3 and 10.4 provided the (equal) lowest NMSE of the yields of total positron annihilation in 5 out of 15 cases. The BIC model in Geant4 10, 10.1 and 10.2 also provided the (equal) lowest NMSE for $^{11}$C fragment production (11/15 cases), whereas for $^{10}$C the best version/model combination was 10.5/INCL (8/15 cases) and for $^{15}$O it was 10.6/BIC (9/12 cases).

\begin{table}
\caption{Number of test cases for which each Geant4 version/model combination achieved a CCC greater than 0.95 in the {\bf entrance region}. \textbf{Bold text} denotes the version/model achieving the highest number of best results for each combination of ion/energy/target.}
\scriptsize
\centering
\begin{tabular}{p{0.08\columnwidth}p{0.04\columnwidth}p{0.04\columnwidth}p{0.04\columnwidth}p{0mm}p{0.04\columnwidth}p{0.04\columnwidth}p{0.04\columnwidth}p{0mm}p{0.04\columnwidth}p{0.04\columnwidth}p{0.04\columnwidth}p{0mm}p{0.04\columnwidth}p{0.04\columnwidth}p{0.04\columnwidth}} 
\toprule 
\multirow{3}{*}{\bf Version}&\multicolumn{3}{c}{\bf Total}&&\multicolumn{3}{c}{\bf $^{11}$C} &&\multicolumn{3}{c}{\bf $^{10}$C} &&\multicolumn{3}{c}{\bf $^{15}$O}\\
\cmidrule{2-4} \cmidrule{6-8} \cmidrule{10-12} \cmidrule{14-16}
&{\bf BIC}	&{\bf QMD} &{\bf INCL}&&{\bf BIC}	&{\bf QMD} &{\bf INCL}&&{\bf BIC}	&{\bf QMD} &{\bf INCL}&&{\bf BIC}	&{\bf QMD} &{\bf INCL}\\
\midrule 
10	    &2	&0	&1	&&\bf 4	&0	&0	&&\bf 2	&\bf 2	&0	&&\bf 3	&0	&\bf 3	\\ \addlinespace[0.5em]
10.1	&1	&0	&1	&&\bf 4	&0	&0	&&\bf 2	&\bf 2	&0	&&\bf 3	&0	&\bf 3	\\ \addlinespace[0.5em]
10.2	&2	&0	&1	&&3	&0	&0	&&\bf 2	&\bf 2	&0	&&\bf 3	&0	&\bf 3	\\ \addlinespace[0.5em]
10.3	&2	&0	&\bf 3	&&3	&1	&3	&&\bf 2	&\bf 2	&0	&&\bf 3	&0	&2	\\ \addlinespace[0.5em]
10.4	&2	&0	&\bf 3	&&\bf 4	&1	&3	&&1	&\bf 2	&0	&&\bf 3	&0	&2	\\ \addlinespace[0.5em]
10.5	&\bf 3	&0	&1	&&3	&2	&\bf 4	&&1	&\bf 2	&0	&&\bf 3	&0	&\bf 3	\\ \addlinespace[0.5em]
10.6	&\bf 3	&0	&1	&&2	&1	&2	&&1	&\bf 2	&0	&&\bf 3	&0	&2	\\ \addlinespace[0.5em]
10.7	&\bf 3	&0	&1	&&3	&1	&3	&&1	&\bf 2	&0	&&\bf 3	&0	&2	\\ \addlinespace[0.5em]
11	    &\bf 3	&0	&1	&&2	&1	&3	&&1	&\bf 2	&0	&&\bf 3	&0	&2	\\ \addlinespace[0.5em]
11.1	&1	&0	&1	&&2	&1	&2	&&1	&\bf 2	&0	&&\bf 3	&0	&2	\\ \addlinespace[0.5em]
\bottomrule
\hline
\end{tabular}
\label{tbl:CCC_Entrance_positron}
\end{table}

Geant4 versions 10.5-11 with BIC and 10.3/10.4 with INCL each achieved a Pearson CCC greater than 0.95 (3/15 cases) for total positron yield; QMD did not reach the threshold for any test case in any version of Geant4.

Results for individual radionuclides were also mixed, with 10/BIC, 10.1/BIC, 10.4/BIC and 10.5/INCL achieving the threshold in 4/15 cases for $^{11}$C, 10-10.4/BIC and all versions with QMD reaching the threshold in 2/15 cases for $^{10}$C, and all versions with BIC and 10/INCL, 10.1/INCL, 10.2/INCL, and 10.5/INCL reaching the threshold for $^{15}$O.

\subsection{Build-up and Bragg peak region}\label{sec:resultsDiscussion-bp}

In the build-up and Bragg peak region, positron-emitting fragments are produced via a combination of target fragmentation and projectile fragmentation. There is a rapid change in positron-emitting fragment yield with respect to depth, especially since different positron-emitting fragments stop at different distances from their point of production.

NMSE and Pearson CCC results between simulation and experimental total positron annihilation profiles in the build-up and Bragg peak region are summarised in Tables \ref{tbl:NMSE_build-upandBraggPeak_positron} and \ref{tbl:CCC_build-upandBraggPeak_positron}, respectively, with corresponding figures shown in Supplementary Material Section 2.

\begin{table}
\caption{Number of test cases for which each Geant4 version/model combination achieved the lowest or equal-lowest NMSE in the {\bf build-up and Bragg peak region}. \textbf{Bold text} denotes the version/model achieving the highest (or equal-highest) number of best results for each combination of ion/energy/target.}
\scriptsize
\centering
\begin{tabular}{p{0.08\columnwidth}p{0.04\columnwidth}p{0.04\columnwidth}p{0.04\columnwidth}p{0mm}p{0.04\columnwidth}p{0.04\columnwidth}p{0.04\columnwidth}p{0mm}p{0.04\columnwidth}p{0.04\columnwidth}p{0.04\columnwidth}p{0mm}p{0.04\columnwidth}p{0.04\columnwidth}p{0.04\columnwidth}} 
\toprule 
\multirow{3}{*}{\bf Version}&\multicolumn{3}{c}{\bf Total}&&\multicolumn{3}{c}{\bf $^{11}$C} &&\multicolumn{3}{c}{\bf $^{10}$C} &&\multicolumn{3}{c}{\bf $^{15}$O}\\
\cmidrule{2-4} \cmidrule{6-8} \cmidrule{10-12} \cmidrule{14-16}
&{\bf BIC}	&{\bf QMD} &{\bf INCL}&&{\bf BIC}	&{\bf QMD} &{\bf INCL}&&{\bf BIC}	&{\bf QMD} &{\bf INCL}&&{\bf BIC}	&{\bf QMD} &{\bf INCL}\\
\midrule 
10	&4	&0	&0	&&11	&6	&3	&&0	&0	&2	&&3	&2	&0	\\ \addlinespace[0.5em]
10.1	&5	&0	&0	&&11	&6	&3	&&0	&0	&2	&&3	&2	&0	\\ \addlinespace[0.5em]
10.2	&\bf 11	&1	&0	&&\bf 14	&6	&3	&&0	&0	&1	&&\bf 5	&1	&0	\\ \addlinespace[0.5em]
10.3	&1	&0	&0	&&5	&0	&0	&&4	&1	&2	&&3	&0	&0	\\ \addlinespace[0.5em]
10.4	&1	&0	&0	&&1	&0	&0	&&3	&2	&2	&&4	&0	&0	\\ \addlinespace[0.5em]
10.5	&0	&0	&0	&&0	&0	&0	&&3	&4	&7	&&1	&2	&0	\\ \addlinespace[0.5em]
10.6	&1	&0	&0	&&0	&0	&0	&&3	&0	&\bf 9	&&2	&1	&2	\\ \addlinespace[0.5em]
10.7	&1	&0	&0	&&0	&0	&0	&&3	&2	&7	&&0	&0	&1	\\ \addlinespace[0.5em]
11	&2	&0	&0	&&0	&0	&0	&&3	&2	&\bf 9	&&2	&0	&2	\\ \addlinespace[0.5em]
11.1	&1	&0	&0	&&0	&0	&0	&&3	&2	&\bf 9	&&2	&0	&2	\\ \addlinespace[0.5em]
\bottomrule
\hline
\end{tabular}
\label{tbl:NMSE_build-upandBraggPeak_positron}
\end{table}

In the build-up and Bragg peak region, according to the NMSE metric, total positron yield is most accurately predicted by the BIC model in Geant4 version 10.2, being (equal) best in 11/15 cases. This is much higher than the next-best combinations (10.1/BIC with 5/11 cases followed by 10/BIC with 4/11). Similar results are observed for $^{11}$C yield, with 10.2/BIC achieving (equal) best performance in 14/15 cases, and 10/BIC and 10.1/BIC each achieving (equal) best results in 11/15 cases; QMD also performs reasonably well in this case with 10, 10.1 and 10.2 achieving wins in 6/15 cases. For $^{10}$C, 10.6/INCL, 11/INCL and 11.1/INCL are the best performers (each winning in 9/15 cases). Finally, for $^{15}$O, 10.2/BIC is the best-performing model with 5/12 wins, followed by 10.4/BIC with 4 wins.

\begin{table}
\caption{Number of test cases for which each Geant4 version/model combination achieved a CCC greater than 0.95 in the {\bf build-up and Bragg peak region}. \textbf{Bold text} denotes the version/model achieving the highest number of best results for each combination of ion/energy/target.}
\scriptsize
\centering
\begin{tabular}{p{0.08\columnwidth}p{0.04\columnwidth}p{0.04\columnwidth}p{0.04\columnwidth}p{0mm}p{0.04\columnwidth}p{0.04\columnwidth}p{0.04\columnwidth}p{0mm}p{0.04\columnwidth}p{0.04\columnwidth}p{0.04\columnwidth}p{0mm}p{0.04\columnwidth}p{0.04\columnwidth}p{0.04\columnwidth}} 
\toprule 
\multirow{3}{*}{\bf Version}&\multicolumn{3}{c}{\bf Total}&&\multicolumn{3}{c}{\bf $^{11}$C} &&\multicolumn{3}{c}{\bf $^{10}$C} &&\multicolumn{3}{c}{\bf $^{15}$O}\\
\cmidrule{2-4} \cmidrule{6-8} \cmidrule{10-12} \cmidrule{14-16}
&{\bf BIC}	&{\bf QMD} &{\bf INCL}&&{\bf BIC}	&{\bf QMD} &{\bf INCL}&&{\bf BIC}	&{\bf QMD} &{\bf INCL}&&{\bf BIC}	&{\bf QMD} &{\bf INCL}\\
\midrule 
10	&9	&4	&3	&&3	&3	&2	&&4	&3	&\bf 6	&&3	&2	&3	\\ \addlinespace[0.5em]
10.1	&9	&4	&4	&&3	&3	&3	&&4	&3	&\bf 6	&&3	&2	&3	\\ \addlinespace[0.5em]
10.2	&\bf 10	&8	&4	&&6	&6	&3	&&5	&4	&5	&&3	&3	&3	\\ \addlinespace[0.5em]
10.3	&9	&6	&6	&&6	&5	&3	&&2	&2	&2	&&\bf 4	&3	&3	\\ \addlinespace[0.5em]
10.4	&6	&8	&7	&&6	&5	&3	&&1	&1	&3	&&\bf 4	&3	&3	\\ \addlinespace[0.5em]
10.5	&8	&8	&6	&&6	&5	&4	&&2	&2	&3	&&3	&3	&\bf 4	\\ \addlinespace[0.5em]
10.6	&9	&9	&6	&&6	&\bf 7	&6	&&2	&1	&4	&&\bf 4	&3	&\bf 4	\\ \addlinespace[0.5em]
10.7	&6	&6	&4	&&3	&5	&2	&&2	&1	&4	&&3	&2	&3	\\ \addlinespace[0.5em]
11	&8	&8	&6	&&4	&5	&4	&&3	&2	&4	&&3	&3	&3	\\ \addlinespace[0.5em]
11.1	&6	&5	&6	&&2	&3	&4	&&3	&2	&4	&&3	&3	&3	\\ \addlinespace[0.5em]
\bottomrule
\hline
\end{tabular}
\label{tbl:CCC_build-upandBraggPeak_positron}
\end{table}

Using the Pearson CCC metric, the best-performing version/model combinations for overall positron yield are 10.2/BIC (10/15 cases), followed by 10/BIC, 10.1/BIC, 10.3/BIC, 10.6/BIC and 10.6/QMD (9/15 cases). Generally, BIC performed very well, with all Geant4 versions achieving (equal) best performance in at least 6 cases. $^{11}$C yield was best predicted by 10.6/QMD (7/15 cases) however many version/model combinations did well here also, with 10.2/BIC, 10.2/QMD, 10.3/BIC, 10.4/BIC, 10.5/BIC, 10.6/BIC and 10.6/INCL all achieving 6/15 wins. $^{10}$C yield was best predicted by 10/INCL and 10.1 INCL (6/15 cases), closely followed by 10.2/BIC and 10.2/INCL which won in 5/15 cases. The best-performing version/model combinations for $^{15}$O yield were 10.3/BIC, 10.4/BIC, 10.5/INCL, 10.6/BIC and 10.6/INCL with 4/15 wins each, and all other version/model combinations achieving 2 or 3 wins.

Table ~\ref{tbl:range_max} lists difference between the experimental and simulation positron peak, while Table~\ref{tbl:range_50max} lists the difference between the 50\% fall off point for the experimental and simulated positron peak.

\begin{table}
\caption{Differences between the depths of the maximum positron annihilation yield in experimental and simulation results. Each voxel has a width of 1.5~mm; the maximum error is always in multiples of 1.5~mm increments.}
\scriptsize
\centering
\begin{tabular}{p{0.08\columnwidth}p{0.08\columnwidth}p{0.08\columnwidth}p{0.10\columnwidth}p{0\columnwidth}p{0.08\columnwidth}p{0.08\columnwidth}p{0.10\columnwidth}p{0\columnwidth}p{0.08\columnwidth}p{0.08\columnwidth}p{0.10\columnwidth}} 
\toprule 
\multirow{3}{*}{\bf Version}&\multicolumn{3}{c}{\bf BIC}    &&\multicolumn{3}{c}{\bf QMD}    &&\multicolumn{3}{c}{\bf INCL}\\
\cmidrule{2-4} \cmidrule{6-8} \cmidrule{10-12}\\
	&{\bf $\mu$ (mm)}	&{\bf $\sigma$ (mm)} &{\bf max (mm)} &&{\bf $\mu$ (mm)}	&{\bf $\sigma$ (mm)} &{\bf max (mm)} &&{\bf $\mu$ (mm)}	&{\bf $\sigma$ (mm)} &{\bf max (mm)}  \\
\midrule 
10 & 1 & 1.85 & 6 && -0.20 & 1.69 & 3 && 1.10 & 3.82 & 10.50 \\ \addlinespace[0.5em] 
10.1 & 1 & 1.85 & 6 && -0.20 & 1.69 & 3 && 1 & 3.91 & 10.50 \\ \addlinespace[0.5em] 
10.2 & 0.60 & 1.37 & 3 && -0.60 & 1.58 & -3 && 0.60 & 3.39 & 9 \\ \addlinespace[0.5em] 
10.3 & 1.30 & 1.78 & 6 && -0.30 & 1.41 & -3 && 2.30 & 4.12 & 9 \\ \addlinespace[0.5em] 
10.4 & 1.60 & 1.55 & 6 && -0.10 & 1.44 & -3 && 4 & 4.45 & 10.50 \\ \addlinespace[0.5em] 
10.5 & 0.69 & 0.99 & 3 && 0.60 & 2.03 & 6 && 2.20 & 4.24 & 9 \\ \addlinespace[0.5em] 
10.6 & 0.60 & 1.37 & 3 && -0.10 & 1.55 & 3 && 0.10 & 2.50 & 7.50 \\ \addlinespace[0.5em] 
10.7 & 1.20 & 1.72 & 4.50 && 0.60 & 1.95 & 4.50 && 0.70 & 3.40 & 10.50 \\ \addlinespace[0.5em] 
11 & 1.10 & 1.65 & 4.50 && 0.60 & 1.95 & 4.50 && 0.60 & 3.09 & 9 \\ \addlinespace[0.5em] 
11.1 & 0.30 & 1.62 & 3 && -0.30 & 1.62 & 3 && 0 & 2.78 & 7.50 \\ \addlinespace[0.5em] 
\bottomrule
\hline
\end{tabular}
\label{tbl:range_max}
\end{table}

\begin{table}
\caption{Differences between the distal depths at which the positron annihilation yield has decreased to 50\% of the peak value in experimental and simulation results. Each voxel has a width of 1.5~mm; the maximum error is always in multiples of 1.5~mm increments.}
\scriptsize
\centering
\begin{tabular}{p{0.08\columnwidth}p{0.08\columnwidth}p{0.08\columnwidth}p{0.10\columnwidth}p{0\columnwidth}p{0.08\columnwidth}p{0.08\columnwidth}p{0.10\columnwidth}p{0\columnwidth}p{0.08\columnwidth}p{0.08\columnwidth}p{0.10\columnwidth}} 
\toprule 
\multirow{3}{*}{\bf Version}&\multicolumn{3}{c}{\bf BIC}    &&\multicolumn{3}{c}{\bf QMD}    &&\multicolumn{3}{c}{\bf INCL}\\
\cmidrule{2-4} \cmidrule{6-8} \cmidrule{10-12}\\
	&{\bf $\mu$ (mm)}	&{\bf $\sigma$ (mm)} &{\bf max (mm)} &&{\bf $\mu$ (mm)}	&{\bf $\sigma$ (mm)} &{\bf max (mm)} &&{\bf $\mu$ (mm)}	&{\bf $\sigma$ (mm)} &{\bf max (mm)}  \\
\midrule
10 & 0.70 & 1.49 & 3 && 0.30 & 1.41 & 3 && -0.20 & 1.49 & -3 \\ \addlinespace[0.5em] 
10.1 & 0.70 & 1.49 & 3 && 0.30 & 1.41 & 3 && -0.20 & 1.49 & -3 \\ \addlinespace[0.5em] 
10.2 & 0.30 & 1.01 & 1.50 && 0 & 1.13 & 1.50 && -0.50 & 1.22 & -3 \\ \addlinespace[0.5em] 
10.3 & 0.50 & 1.09 & 3 && 0.20 & 1.11 & 1.50 && -0.20 & 1.37 & -3 \\ \addlinespace[0.5em] 
10.4 & 0.60 & 1.11 & 3 && 0.30 & 1.01 & 1.50 && 0.10 & 1.20 & -3 \\ \addlinespace[0.5em] 
10.5 & 0.35 & 0.90 & 1.50 && 0.20 & 1.11 & 1.50 && -0.50 & 1.22 & -3 \\ \addlinespace[0.5em] 
10.6 & 0.40 & 0.89 & 1.50 && 0.20 & 1.11 & 1.50 && -0.40 & 1.33 & -3 \\ \addlinespace[0.5em] 
10.7 & 1 & 1.46 & 3 && 0.70 & 1.59 & 3 && -0.10 & 1.65 & 3 \\ \addlinespace[0.5em] 
11 & 1 & 1.46 & 3 && 0.70 & 1.59 & 3 && 0 & 1.60 & 3 \\ \addlinespace[0.5em] 
11.1 & 0 & 1.60 & -3 && -0.10 & 1.44 & -3 && -0.60 & 1.68 & -3 \\ \addlinespace[0.5em]   
\bottomrule
\hline
\end{tabular}
\label{tbl:range_50max}
\end{table}

The smallest differences between experimental and simulation-based depth of maximum positron annihilation were obtained with Geant4 10.4/QMD ($\mu = -0.1$~mm; max = -3~mm) and 10.6/QMD ($\mu = -0.1$~mm; max = +3~mm). While a smaller mean value was obtained with 11.1/INCL, the maximum value and standard deviation were much larger (+7.5~mm and 2.78~mm) compared to 10.4/QMD and 10.6/QMD. Differences in the depth of the distal 50\%-of-peak point were much smaller; the best estimates were obtained with 10.2/QMD ($\mu = 0$~mm; max = +1.5~mm), 11.1/BIC ($\mu = 0$~mm; max = -3~mm) and 11/INCL ($\mu = 0$~mm; max = +3~mm).

\subsection{Tail region}\label{sec:resultsDiscussion-tail}

In the tail region, positron-emitting radionuclides are primarily produced through fragmentation of the target material caused by light fragments created upstream from the primary beam. As such, the production of positron-emitting fragments in the tail region is highly dependent on fragmentation and scattering cross sections upstream. Therefore, the yield of positron annihilation is not expected to rapidly change across this region compared to the build-up and Bragg peak region.

NMSE and Pearson CCC results between simulation and experimental total positron annihilation profiles in the tail region are summarised in Tables \ref{tbl:NMSE_tail_positron} and \ref{tbl:CCC_tail_positron}, respectively, with corresponding figures shown in Supplementary Material Section 3.

\begin{table}
\caption{Number of test cases for which each Geant4 version/model combination achieved the lowest or equal-lowest NMSE in the {\bf tail region}. \textbf{Bold text} denotes the version/model achieving the highest (or equal-highest) number of best results for each combination of ion/energy/target.}
\scriptsize
\centering
\begin{tabular}{p{0.08\columnwidth}p{0.04\columnwidth}p{0.04\columnwidth}p{0.04\columnwidth}p{0mm}p{0.04\columnwidth}p{0.04\columnwidth}p{0.04\columnwidth}p{0mm}p{0.04\columnwidth}p{0.04\columnwidth}p{0.04\columnwidth}p{0mm}p{0.04\columnwidth}p{0.04\columnwidth}p{0.04\columnwidth}} 
\toprule 
\multirow{3}{*}{\bf Version}&\multicolumn{3}{c}{\bf Total}&&\multicolumn{3}{c}{\bf $^{11}$C} &&\multicolumn{3}{c}{\bf $^{10}$C} &&\multicolumn{3}{c}{\bf $^{15}$O}\\
\cmidrule{2-4} \cmidrule{6-8} \cmidrule{10-12} \cmidrule{14-16}
&{\bf BIC}	&{\bf QMD} &{\bf INCL}&&{\bf BIC}	&{\bf QMD} &{\bf INCL}&&{\bf BIC}	&{\bf QMD} &{\bf INCL}&&{\bf BIC}	&{\bf QMD} &{\bf INCL}\\
\midrule 
10	&3	&2	&2	&&5	&9	&4	&&2	&2	&1	&&4	&4	&4	\\ \addlinespace[0.5em]
10.1	&4	&2	&2	&&6	&9	&4	&&1	&1	&2	&&4	&4	&4	\\ \addlinespace[0.5em]
10.2	&\bf 12	&10	&2	&&11	&\bf 12	&4	&&1	&1	&3	&&7	&7	&4	\\ \addlinespace[0.5em]
10.3	&1	&1	&1	&&1	&1	&1	&&2	&4	&0	&&3	&4	&4	\\ \addlinespace[0.5em]
10.4	&1	&1	&1	&&1	&1	&1	&&2	&3	&0	&&3	&4	&3	\\ \addlinespace[0.5em]
10.5	&1	&1	&1	&&1	&1	&1	&&3	&4	&2	&&2	&2	&2	\\ \addlinespace[0.5em]
10.6	&1	&2	&1	&&1	&1	&1	&&5	&3	&\bf 6	&&4	&\bf 10	&4	\\ \addlinespace[0.5em]
10.7	&1	&1	&1	&&1	&1	&1	&&3	&3	&5	&&3	&4	&3	\\ \addlinespace[0.5em]
11	&1	&1	&1	&&1	&1	&1	&&4	&3	&\bf 6	&&4	&4	&3	\\ \addlinespace[0.5em]
11.1	&1	&1	&1	&&1	&1	&1	&&4	&3	&5	&&4	&4	&3	\\ \addlinespace[0.5em]
\bottomrule
\hline
\end{tabular}
\label{tbl:NMSE_tail_positron}
\end{table}

Using the NMSE metric, 10.2/BIC was the best-performing version/model combination for overall positron yield (12/15 cases), with 10.2/QMD being the second-best (10/15). Results were similar for $^{11}$C yield, with the best version/model combinations being 10.2/QMD (12/15 cases) and 10.2/BIC (11/15). For $^{10}$C, the most wins were obtained by 10.6/INCL and 11/INCL (6/15 cases) followed by 10.6/BIC, 10.7/INCL and 11.1/INCL (5/15 cases). Finally, for $^{15}$O, the best results were obtained with 10.6/QMD (10/12 cases) followed by 10.2/BIC and 10.2/QMD (7/12 cases).

\begin{table}
\caption{Number of test cases for which each Geant4 version/model combination achieved a CCC greater than 0.95 in the {\bf tail region}. \textbf{Bold text} denotes the version/model achieving the highest number of best results for each combination of ion/energy/target.}
\scriptsize
\centering
\begin{tabular}{p{0.08\columnwidth}p{0.04\columnwidth}p{0.04\columnwidth}p{0.04\columnwidth}p{0mm}p{0.04\columnwidth}p{0.04\columnwidth}p{0.04\columnwidth}p{0mm}p{0.04\columnwidth}p{0.04\columnwidth}p{0.04\columnwidth}p{0mm}p{0.04\columnwidth}p{0.04\columnwidth}p{0.04\columnwidth}} 
\toprule 
\multirow{3}{*}{\bf Version}&\multicolumn{3}{c}{\bf Total}&&\multicolumn{3}{c}{\bf $^{11}$C} &&\multicolumn{3}{c}{\bf $^{10}$C} &&\multicolumn{3}{c}{\bf $^{15}$O}\\
\cmidrule{2-4} \cmidrule{6-8} \cmidrule{10-12} \cmidrule{14-16}
&{\bf BIC}	&{\bf QMD} &{\bf INCL}&&{\bf BIC}	&{\bf QMD} &{\bf INCL}&&{\bf BIC}	&{\bf QMD} &{\bf INCL}&&{\bf BIC}	&{\bf QMD} &{\bf INCL}\\
\midrule 
10	&\bf 12	&11	&11	&&10	&11	&11	&&\bf 4	&3	&\bf 4	&&8	&7	&\bf 9	\\ \addlinespace[0.5em]
10.1	&\bf 12	&11	&11	&&11	&11	&11	&&\bf 4	&\bf 4	&\bf 4	&&8	&7	&\bf 9	\\ \addlinespace[0.5em]
10.2	&\bf 12	&11	&11	&&10	&11	&11	&&\bf 4	&\bf 4	&\bf 4	&&8	&7	&7	\\ \addlinespace[0.5em]
10.3	&11	&11	&10	&&11	&11	&11	&&2	&2	&1	&&8	&7	&8	\\ \addlinespace[0.5em]
10.4	&10	&10	&10	&&9	&11	&10	&&2	&2	&0	&&7	&7	&8	\\ \addlinespace[0.5em]
10.5	&11	&11	&11	&&11	&11	&11	&&2	&2	&3	&&7	&7	&8	\\ \addlinespace[0.5em]
10.6	&11	&11	&\bf 12	&&11	&11	&11	&&2	&3	&3	&&7	&7	&\bf 9	\\ \addlinespace[0.5em]
10.7	&\bf 12	&11	&\bf 12	&&11	&11	&12	&&3	&3	&3	&&7	&7	&\bf 9	\\ \addlinespace[0.5em]
11	&\bf 12	&11	&\bf 12	&&11	&11	&\bf 13	&&3	&3	&3	&&8	&8	&\bf 9	\\ \addlinespace[0.5em]
11.1	&\bf 12	&11	&\bf 12	&&11	&11	&\bf 13	&&3	&3	&3	&&8	&8	&\bf 9	\\ \addlinespace[0.5em]
\bottomrule
\hline
\end{tabular}
\label{tbl:CCC_tail_positron}
\end{table}

The Pearson CCC results in the tail region were all very similar across Geant4 versions, with only a few wins separating the best and worst-performing version/model combinations in most instances. All version/models exceeded the threshold of 0.95 for a clear majority of cases for total positron yield as well as $^{11}$C and $^{15}$O production. For total positron annihilation yield, 10/BIC, 10.1/BIC, 10.2/BIC, 10.6/INCL, 10.7/BIC, 10.7/INCL, 11/BIC, 11/INCL, 11.1/BIC and 11.1/INCL all exceeded the target threshold for 12/15 cases. Even the worst-performing version/model combinations still exceeded the threshold in 10/15 cases. For $^{11}$C yield, 11/INCL and 11.1/INCL reached the threshold in 13/15 cases (with the worst-performing combination scoring 9/15 wins). Fewer wins were seen with $^{10}$C; the best results were obtained with 10/BIC, 10/INCL, 10.1/BIC, 10.1/QMD, 10.1/INCL, 10.2/BIC, 10.2/QMD and 10.2/INCL (4/15 cases). Finally, $^{15}$O yield was best predicted by 10/INCL, 10.1/INCL, 10.6/INCL, 10.7/INCL, 11/INCL and 11.1/INCL (9/12 cases) - again, in this case, even the worst-performing version/model combinations exceeded the threshold in 7/12 cases.

\subsection{Overall recommendation}\label{sec:resultsDiscussion-overall}

The accuracy of Geant4's hadronic inelastic physics models (BIC, QMD and INCL) in predicting both total positron annihilation yield and individual positron-emitting fragment production is not consistent between different versions of Geant4; furthermore, later releases do not necessarily provide a more accurate prediction of experimental observations than preceding versions.

In the entrance region, BIC was clearly the best-performing model, with the best choice of Geant4 version depending on the particular metric and fragmentation product of interest. NMSE results generally favoured 10-10.4/BIC (especially 10.2/BIC), except for $^{10}$C yield, which was better predicted by 10.3+/INCL. Pearson CCC performance did not strongly favour any particular version/model combination, with at most 1/3 of test cases achieving the target CCC threshold of 0.95 for any version/model.

In the build-up and Bragg peak region and tail region, the results are more clear-cut. Results for the NMSE metric conclusively show that version 10.2/BIC is clearly the best choice for total positron yield, $^{11}$C production and $^{15}$O yield, while 10.5-11.1/INCL performed the best for $^{10}$C. Pearson CCC results are more mixed, but again, 10.2/BIC gives the best results for total positron annihilation yield, with most versions of Geant4 with BIC performing well. 10.6/QMD performed the best for $^{11}$C, 10/INCL and 10.1/INCL performed the best for $^{10}$C, and there was no clear winner for $^{15}$O.

Using the depth-of-maximum-yield metric, the smallest mean differences were obtained with 10.4/QMD and 10.6/QMD. These versions/models also achieved the equal-smallest maximum difference (-3~mm and +3~mm, respectively). Across all versions of Geant4, QMD demonstrated the best overall accuracy (lowest average mean difference in peak depth) and highest precision (lowest average standard deviation). INCL was the worst-performing model across all versions, with much larger maximum differences, and a consistent underestimation of depth of maximum yield across Geant4 versions, with the exception of version 11.1 (which, despite a mean difference of 0, exhibited a large standard deviation and maximum value). Standard deviations obtained using INCL were generally around double those of QMD and BIC. BIC also showed a consistent underestimation in depth of maximum yield, although the maximum differences were much smaller than for INCL. For context, the difference between the depth of the positron annihilation peak and the Bragg peak with monoenergetic ion beams is of the order of -5.6$\pm$0.8~mm for $^{12}$C and -6.6$\pm$0.8~mm for $^{16}$O \cite{Chacon2020,Augusto2018,Mohammadi_2019}.

Results were generally better for the the distal depth at 50\% of peak metric. In this case, 10.2/QMD, 11.1/BIC and 11/INCL all achieved a mean of zero, with 10.2/QMD also having the equal-lowest maximum value of 1.5~mm (a depth difference of one voxel). QMD's maximal values were slightly smaller overall compared to BIC, and INCL's were the largest at $\pm$3~mm for all versions. INCL tended to consistently overestimate the depth of this point, with both mean and maximum differences being negative in most cases. BIC and QMD both tended towards underestimating the 50\%-of-peak depth, with the exception of version 11.1 (negative maxima for both, and means of 0 and -0.1~mm, respectively). Standard deviations were quite small for all versions and models (with the maximum standard deviation being 1.68~mm, for 11.1/INCL).



Finally, in the tail region, Geant4 10.2 with BIC and QMD again provided the best prediction of total positron and $^{11}$C yield in terms of NMSE, while 10.6/INCL performed the best for $^{10}$C and $^{15}$O. All version/model combinations performed well for total positron annihilation, $^{11}$C and $^{15}$O yield according to the Pearson CCC metric, while no version/model performed especially well for $^{10}$C.

Across all regions, ion species, beam energies, and target materials evaluated, the combination of Geant4 version 10.2 and BIC is best able to reproduce experimental results as evaluated using the NMSE and Pearson CCC metrics - especially in the build-up and Bragg peak region and tail region. Since the build-up and Bragg peak region is the location where (1) the majority of the dose resulting from carbon or oxygen ion beam irradiation in heavy ion therapy is deposited, and (2) where the strongest positron annihilation signal is observed, the results in this region are the most relevant to PET image-based QA simulation work. Version 10.2 also provided the best estimate of the depth of the distal depth at which positron yield decreased to 50\% of peak, although this was obtained with QMD rather than BIC; the most accurate estimate of the depth of the peak itself was also achieved with QMD, but with Geant4 versions 10.4 and 10.6. As QMD exhibited the best accuracy and precision across Geant4 versions, it is the recommended model if the depth of the yield peak is critical.

The BIC model implemented in Geant4 version 10.5 suffered from a run-time stability error which resulted in it being unable to simulate all test scenarios; therefore, we recommend that this version/model combination should be avoided for future studies. 

In the evaluation of individual positron-emitting fragment yield profiles, predictions of $^{10}$C distribution were generally the least accurate in terms of both the NMSE and Pearson CCC. Interestingly, the INCL model often performed the best for prediction of $^{10}$C fragment yield, although it rarely performed the best for total positron annihilation and $^{11}$C or $^{15}$O. Therefore, INCL should be considered for studies focusing on $^{10}$C fragmentation, with the caveat that range estimation will be less accurate with this model.

Not all models met or exceeded the set threshold of 0.95 for the Pearson CCC metric. This means that in these cases, the shape of the predicted positron distribution differs significantly from the experimental measurements. This is of particular concern if these models are to be used for dose estimation using a deconvolution approach \cite{2019Hofmanna,2019Hofmannb} or for the training of machine learning models for feature extraction \cite{Rutherford2022}.


Finally, it is worth noting that current evaluations of fragmentation cross sections exhibit uncertainties exceeding 10\%, which must be addressed in order to accurately model positron fragmentation, particularly in the case of complex fragmentation reactions such as the production of $^{10}$C \cite{Bolst2017,Toppi2022}. This also impacts other Monte Carlo simulation platforms (such as FLUKA, MCNP and PHITS) which rely on accurate cross section data.

\section{Conclusion}\label{sec:conclusion}

In this study, the accuracy with which Geant4 is able to predict the distribution of total positron annihilation yield and the distributions of individual positron-emitting fragmentation products ($^{11}$C, $^{10}$C and $^{15}$O) during carbon or oxygen ion therapy was compared to experimental data. Three different hadronic inelastic physics models - Binary Ion Cascade (BIC), Quantum Molecular Dynamics (QMD) and Liege Intranuclear Cascade model (INCL) were used with ten different versions of Geant4 - 10.0.p04, 10.1.p03, 10.2.p03, 10.3.p03, 10.4.p03, 10.5.p01, 10.6.p03, 10.7.p02, 11.0 and 11.1, in three different homogeneous phantoms. The simulated and experimental data were compared using two different metrics - normalised mean squared error and the Pearson cross-correlation coefficient. Additionally, the differences between the simulated and experimental depth of maximum positron annihilation yield, as well as the distal point at which positron yield declines to 50\% of the peak were evaluated. It was found that the accuracy of the hadronic inelastic physics models strongly depends on the version of Geant4 in which it was implemented, and newer versions of Geant4 were not always more accurate at predicting positron-emitting fragmentation compared to older versions. Furthermore, it was found that not all version/model combinations were able to satisfactorily predict the shape of positron annihilation or positron-emitting fragment distributions, even though they could provide a good estimation of the total positron annihilation yield and range. For future simulation studies of therapeutic irradiation using carbon or oxygen ion beams, it is recommended that Geant4 version 10.2 with the BIC model be used as it is currently the version/model combination best able to replicate the experimentally-observed total positron yield and the fragmentation product distributions, while the depth of the maximum positron yield and distal 50\%-of-peak point were best predicted using the QMD model from Geant4 10.4, 10.6 (peak) and 10.2 (distal 50\%-of-peak).

\section{Acknowledgements}

The authors would like to acknowledge the following organisations for providing access to their high-performance computing resources: the Multi-modal Australian Sciences Imaging and Visualisation Environment (MASSIVE) ``M3'' cluster and Australia's Nuclear Science and Technology Organisation (ANSTO) ``Tesla'' cluster. This research has been conducted with the support of the Australian government research training program scholarship. The authors acknowledge the scientific and technical assistance of the National Imaging Facility, a National Collaborative Research Infrastructure Strategy (NCRIS) capability at the Australian Nuclear Science and Technology Organisation, ANSTO.

\Urlmuskip=0mu plus 1mu

\section*{References}
\bibliography{bibliography}

\section{Appendix}\label{sec:app}

Table ~\ref{tbl:physics-list} lists the physics models which were used in the simulations.

\begin{table}[!htb]
	\centering
	\caption{Hadronic physics processes and models used in all simulations.}
	\begin{tabular}{p{0.29\columnwidth}p{0.29\columnwidth}p{0.29\columnwidth}}
		\toprule
		{\bf Interaction}	&{\bf Energy Range }	&{\bf Geant4 Model }\\
		\midrule
		Radioactive Decay	&All energies			&G4RadioactiveDecayPhysics\\
		\addlinespace[0.5em]
		Particle Decay		&All energies			&G4Decay\\
		\addlinespace[0.5em]
		Hadron Elastic		&0--100~TeV		&G4HadronElasticPhysicsHP\\
		\addlinespace[0.5em]
		Ion Inelastic		&$<$100~MeV		&Binary Light Ion Cascade\\
		&100~MeV--10~GeV	&BIC or QMD or INCL++\\
		\addlinespace[0.5em]
		Neutron Capture		&0--20~MeV		&NeutronHPCapture\\
		&$>$19.9~MeV	&nRadCapture\\
		\addlinespace[0.5em]
		Neutron Inelastic	&0--20~MeV		&NeutronHPInelastic\\
		&$>$19.9~MeV	&Binary Cascade\\
		\addlinespace[0.5em]
		Proton Inelastic	& 990~eV--10~TeV		&Binary Cascade\\
		\bottomrule
	\end{tabular}
	\label{tbl:physics-list}
\end{table}




\end{document}